\documentclass[aps,twocolumn,preprintnumbers,amsmath,amssymb]{revtex4-2}

\usepackage{graphicx}
\usepackage{dcolumn}
\usepackage{bm}
\usepackage{multirow}
\usepackage{color}
\usepackage{ulem}
\usepackage{amsmath}
\usepackage{gensymb}

\begin{document}
\title{Enhancement of valley  polarization  at high photoexcited densities in MoS$_2$ monolayers.}

\email{fabian.cadiz@polytechnique.edu}
\author{F. Cadiz$^{1}$}
\author{S. Gerl $^{1}$}
\author{T. Taniguchi$^2$}
\author{K. Watanabe$^3$}

\affiliation{$^1$ Laboratoire de Physique de la Mati\` ere Condens\' ee, CNRS, Ecole Polytechnique, Institut Polytechnique de Paris, 91120 Palaiseau, France}

\affiliation{$^2$International Center for Materials Nanoarchitectonics, National Institute for Materials Science, 1-1 Namiki, Tsukuba 305-0044, Japan}

\affiliation{$^3$ Research Center for Functional Materials, National Institute for Materials Science, 1-1 Namiki, Tsukuba 305-0044, Japan}


\begin{abstract}
We have investigated the steady-sate valley polarization and valley coherence of encapsulated MoS$_2$ monolayer as a function of the temperature and the power density with a continuous wave laser excitation. Both valley polarization and coherence exhibit a non-monotonic dependence on sample temperature, attaining a local maximum at $T\approx 40$ K. This has been recently attributed to a motional narrowing effect: an enhancement of the valley relaxation time occurs when  the scattering rate increases.
At a fixed temperature of $T=6$ K, a two-fold increase of the steady-state valley polarization is achieved by increasing the laser excitation power, which we attribute to a local heating induced by the energy relaxation of photoexcited excitons outside the light cone and to an increase in the exciton-exciton scattering rate. In contrast, in the same power range only a moderate enhancement of valley coherence is observed. Further increasing  the excitation power leads to a small reduction of valley polarization but a dramatic loss of valley coherence.  Supported by spatial imaging of the excitonic luminescence and polarization,  we attribute this behaviour to the detrimental role of exciton-exciton interactions on the pure dephasing rate.
\end{abstract}


\maketitle
\textit{Introduction.---}
Atomically-thin layers of  transition metal dichalcogenides (TMD) such as MX$_2$ (M=Mo, W; X=S, Se, Te) have emerged as promising 2D semiconductors for applications in valley/spintronics \cite{behnia:2012,Xiao:2012a,Butler:2013a}.  In monolayers, the interplay between inversion symmetry breaking and the strong spin-orbit interaction inherent to the heavy transition metal atoms yields a unique spin/valley texture at the $K^+$/$K^-$ points of the Brillouin zone, which is expected to provide additional functionalities in future devices \cite{Xiao:2012a,Li:2020a,Li:2020b,Huang:2020a}.  Due to enhanced Coulomb interaction in 2D, weak dielectric screening and large effective masses, the optical excitation couples mostly to exciton resonances \cite{Ramasubramaniam:2012a,Ross:2013a,Mak:2013a}. Remarkably, light absorption due to these strongly bound excitons preserves the single-particle coupling between light chirality and the valley degree of freedom \cite{Cao:2012a,Mak:2012a,Zeng:2012a,Sallen:2012a,Kioseoglou:2012a}. Moreover, the short exciton lifetime at cryogenic temperatures \cite{Robert:2016a} is comparable to the dephasing time \cite{Jakubczyk:2016a}, so that coherent superpositions of $K^+$ and $K^-$ valley excitons can be detected in simple steady-state photoluminescence (PL) experiments \cite{Jones:2013a,Wang:2017a,Cadiz:2017a}. However, there is still a lack of understanding of the different microscopic mechanisms that govern valley polarization and coherence in TMD monolayers. In this work, we have fabricated encapsulated monolayer MoS$_2$ heterostructures; and we present an investigation of the steady-state valley polarization and coherence of neutral excitons as a function of sample temperature and excitation power density. At low excitation power, we find that both valley polarization and coherence attain a local maximum at $40$ K, in agreement with very recent observations \cite{Wu:2021a}.  We show that, at a fixed temperature of $T=6$ K,  a similar enhancement of the valley polarization can be achieved by increasing the laser power density. The valley polarization attains a maximum at a photoexcited density of $n^* \sim 10^{10}\; \mbox{cm}^{-2}$, and slowly decreases for densities above $n^*$. We attribute this to two laser-related effects: local heating and increased rate of exciton-exciton scattering events. The latter is particularly detrimental to valley coherence, for which we see only a small enhancement at $n^*$, and a dramatic decrease upon further increase of the excitation power. Spatially-resolved PL shows that valley coherence develops a significant spatial gradient at high densities, reflecting the important role of the exciton density on the valley dynamics. \\

\begin{figure*}
\includegraphics[width=1.05\textwidth]{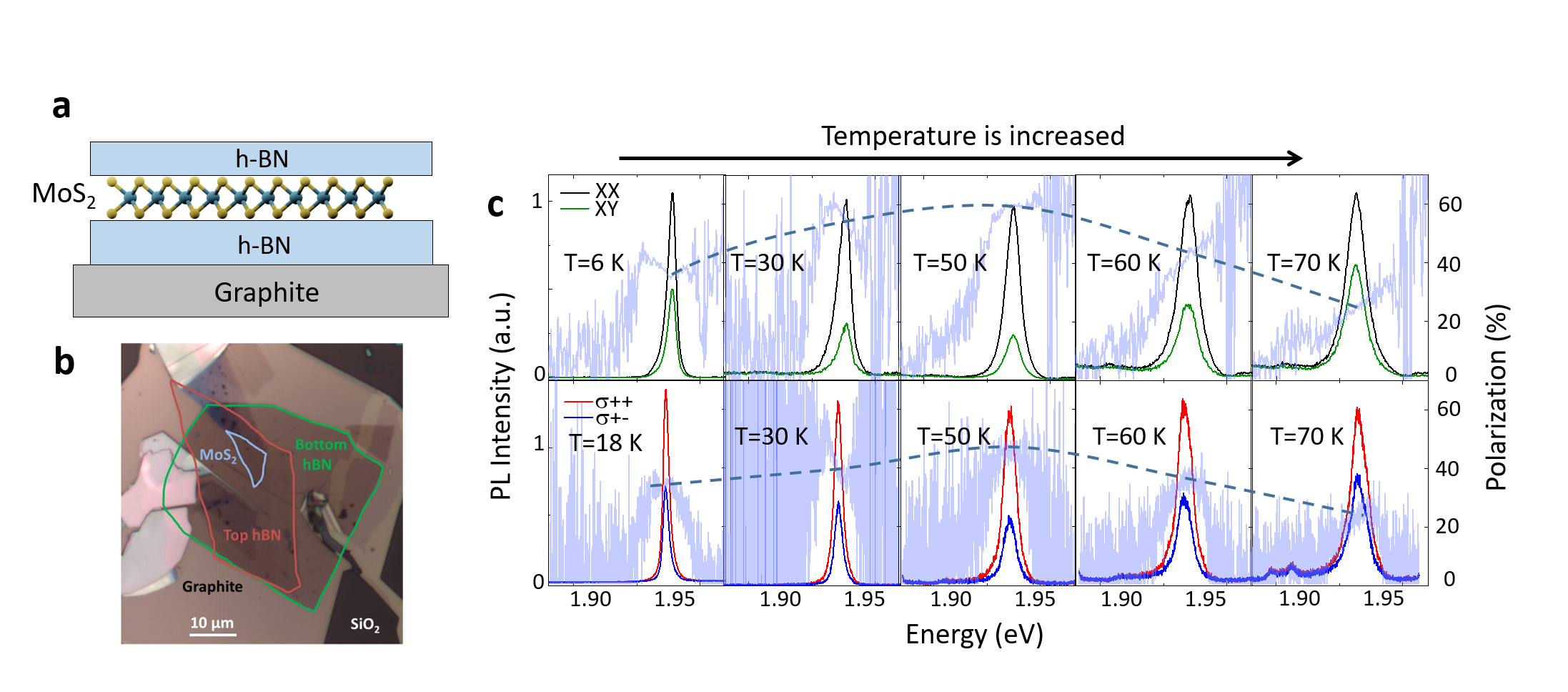}
\caption{\label{fig:fig1} (a) Schematic side-view of the sample. The MoS$_2$ monolayer is encapsulated between two thin h-BN flakes to provide high optical quality and to prevent photodoping effects. A thin graphite flake is used to screen disorder from the substrate and to avoid back reflections for PL imaging. The whole heterostructure is deposited onto a silicon substrate with a 90 nm-thick silicon dioxide  layer. (b) Microscope image of the sample. (c) Polarization-resolved PL spectra under linear (top) and circular (bottom) excitation at 1.96 eV for different sample temperatures. The excitation power was kept at 20 $\mu$W. Also shown is the degree of polarization as a function of photon emission energy. The dashed lines are a guide to the eye indicating the polarization at the energy at with the PL intensity is maximum.}
\end{figure*}

\indent \textit{Samples and Experimental Set-up.---}
Encapsulated MoS$_2$ monolayers such as the one shown in Fig.\ref{fig:fig1} (a)-(b) were fabricated by mechanical exfoliation of bulk molybdenite crystals from 2D semiconductors. The layers were deterministically and sequentially transferred onto an SiO$_2$ (90 nm)/Si substrate by using a transparent viscoelastic stamp \cite{Gomez:2014a}. 
A hyperspectral confocal  micro-PL  set-up is used to excite and detect the polarized exciton emission at cryogenic temperatures \cite{Favorskiy:2010a, Cadiz:2018a}. The samples are excited with a continuous wave (cw) solid-state laser at 633 nm ($\sim 1.96$ eV), tightly focused onto a diffraction-limited spot in the sample plane. At $T=6$ K, this laser is detuned by 23 meV from the neutral exciton transition. The Airy-disk of the laser's intensity on the sample plane can be approximated by a gaussian profile of the form $e^{-r^2/\sigma^2}$, with $r$ the radial distance from the center of the laser spot and $\sigma\approx 0.3\;\mu$m. The polarization of both the laser and the detected PL is controlled with liquid crystal retarders and linear polarizers. 
  The resulting PL spot is imaged onto the entrance slit of a $320$ mm focal length spectrometer 
 equipped with a 600 grooves/mm diffraction grating.
 For  PL imaging, tuneable filters were used to select the neutral exciton emission, whose spatial distribution was then imaged onto a cooled Si-CCD camera.  \\

\indent \textit{Results and Discussion.---}
Fig.\ref{fig:fig1} (a) shows a schematic drawing of the MoS$_2$-based Van der Waals heterostructure deposited onto a SiO$_2$/Si substrate. A thin graphite flake is used to screen possible charge puddles located on the SiO$_2$ substrate \cite{Barbone:2018a} and also to minimize reflections of the PL coming from the substrate, which could affect PL imaging. A microscope image of the sample under white light illumination is shown in Fig.\ref{fig:fig1}(b). The high-optical quality of our sample is confirmed by a $\sim 2$ meV neutral exciton linewith in PL at low temperatures and low excitation power, close to the homogeneous limit \cite{Cadiz:2017a,Jakubczyk:2016a,Martin:2020a}.\\

\begin{figure*}[htbp]
\includegraphics[width=1.05\textwidth]{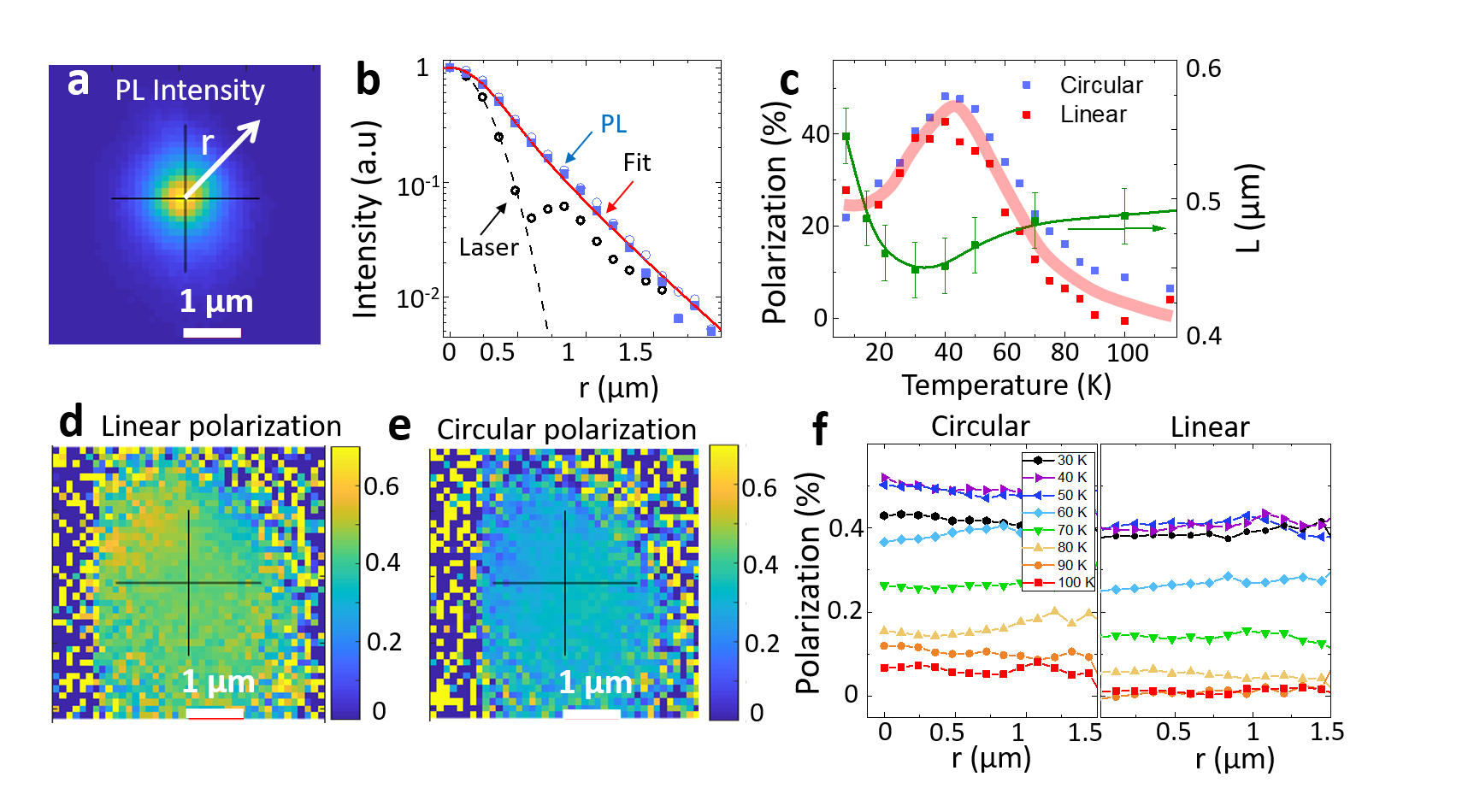}
\caption{\label{fig:fig2} (a) Spatially-resolved PL intensity under cw laser excitation at $5\;\mu$W at T=6 K. The radial distance with respect to the excitation spot is denoted by $r$. (b) Radial profile of the PL intensity shown in (a) obtained after averaging over different directions, for a circular (squares) and linear (open circles) laser excitation. Also shown is the laser profile (black circles) together with a Gaussian profile (dashed black line) with a radius of $0.3\; \mu$m. The red line is a fit with a solution of the 2D diffusion equation (Eq.(\ref{eq1})) giving an effective diffusion length of $L=0.52 \pm 0.025\;\mu$m.  (c) Degree of polarization under circular and linear excitation for an excitation power of 20 $\mu$W as a function of sample temperature. Also shown is the effective diffusion length L extracted from the spatially-resolved PL images. Continuous lines are a guide to the eye. (d) Spatially-resolved linear polarization under linear excitation for the same conditions as (a). (e) Spatially-resolved circular polarization under circular excitation for the same conditions as (a). (f) Radial profile of the photoluminescence's  circular and linear polarization for selected values of the sample temperature. The laser excitation is kept at 20 $\;\mu$W. }
\end{figure*}

\noindent
Figure \ref{fig:fig1}(c) shows the spatially-averaged PL spectrum, decomposed into its co-polarized ($I_{\mbox{co}}$) and cross-polarized  ($I_{\mbox{cross}}$) components with respect to the laser polarization for a low excitation power of $20\;\mu$W. Also shown is the degree of polarization at each emitted photon energy, defined as $\mathcal{P}= (I_{\mbox{co}}-I_{\mbox{cross}})/(I_{\mbox{co}}+I_{\mbox{cross}})$. Under circular excitation, excitons are selectively generated in either the $K^+$ or the $K^-$ valley depending on the laser's helicity. The steady-state degree of circular polarization of the PL, smaller than $100\;\%$, reflects valley depolarization during the exciton lifetime.  Under linear excitation, a coherent superposition of excitons in the $K^+$ and $K^-$ valley is generated \cite{Jones:2013a}. Due to the very short exciton lifetime, in the picosecond (ps) range \cite{Zhu:2014a,Yan:2015b,Wang:2014b,Glazov:2015a,Robert:2016a}, this valley coherence is partially preserved before radiative recombination. Remarkably, increasing the sample temperature  leads to an increase of both the valley polarization and the valley coherence, both peaking at $T\sim 40$ K, before decreasing again upon further increase of the temperature. This is in agreement with very recent findings \cite{Wu:2021a}. The degree of valley polarization can reach $40\;\%$ at $T=40$ K,  whereas valley coherence can be as high as $60\;\%$ although this maximum value was found to be sample and position-dependent, probably due to inhomogeneities of the dielectric environment. The decrease of valley polarization and coherence at higher temperatures can be due to ultrafast intervalley relaxation driven by phonon mediated processes, which becomes faster than 1 ps above 100 K \cite{Wang:2018a}.

 This non-monotonic temperature-dependence was attributed by Wu et al. to a scattering-induced enhancement of valley polarization, relevant when the latter is limited by the long-range electron-hole exchange interaction,  in the so-called Maialle-Silve-Sham (MSS) mechanism\cite{Zhu:2014a}. Indeed, this interaction is equivalent to an effective magnetic field around which the valley pseudospin precesses with a Larmor frequency $\Omega(\vec k)$, where $\vec k$ is the exciton's center-of-mass momentum. Thermally-activated momentum scattering supresses valley pseudospin relaxation whenever the momentum relaxation time $\tau_c$ becomes much shorter than the pseudospin precession time $\Omega^{-1}$, similar to the D'yakonov-Perel spin-relaxation mechanism in non-centrosymmetric semiconductors \cite{Dyakonov:1972a}. Under a tightly-focused laser excitation, we can therefore expect the valley dynamics to be strongly power and spatially-dependent, which is why we will now focus on PL imaging at different photoexcited exciton densities. \\

\noindent
Figure \ref{fig:fig2}(a) shows the resulting image of the exciton luminescence at $T=6$ K and $5\;\mu$W  excitation power. The PL spatial distribution exhibits a clear rotational symmetry,  so that the PL intensity depends only on the distance $r$ with respect to the laser spot. A radial profile of the PL is obtained by averaging cuts along different directions, the result for Fig.\ref{fig:fig2}(a) is shown in Fig.\ref{fig:fig2}(b) for both circular and linear excitation (blue squares and open circles, respectively).

 Also shown is the normalized radial profile of the laser (black open circles), and its Gaussian fit (dashed line). Since the PL clearly extends beyond the laser spot, we can obtain an effective diffusion length $L$ by fitting the PL intensity $I$ with a solution of the steady-state diffusion equation in 2D:

\begin{equation}
I(r) \propto \int_{-\infty}^{+\infty} K_0\left(|r|/L\right) e^{-(r-r')^2/\sigma^2}   dr'
\label{eq1}
\end{equation}

\noindent
where $K_0$ is the modified Bessel function of the second kind and $\sigma=0.3\;\mu$m. This fit is shown by a red line in Fig.\ref{fig:fig2}(b) and yields a diffusion length of $L=0.55 \pm 0.025 \;\mu$m at $T=6$ K.  Such a large diffusion length is unlikely to reflect diffusion of bright excitons. Indeed, considering a population decay time of $\tau\sim 5$ ps \cite{Robert:2016a}, an exciton mass of $m_X \sim m_0$, where $m_0$ is the electron mass,  a temperature of $T=10$ K and a momentum relaxation time of $\tau_c=0.05-1$ ps \cite{Wu:2021a}, we expect an exciton diffusion length in the range \begin{equation} L_X= \sqrt{ \frac{k_B T \tau_c}{ m_X}  \tau  }\approx 6 - 28 \; \mbox{nm}
 \end{equation}
where $k_B$ is Boltzmann's constant (here we have assumed that Einstein's relation is valid even for such a short-lived species). This is at least 18 times smaller than the observed effective diffusion length of $L\approx 0.5\;\mu$m at $T=6$ K.  Moreover, it has been predicted that exciton transport should be anisotropic under linear excitation \cite{Ghazaryan:2018a}, but as shown in Fig.\ref{fig:fig2}(b), no difference is observed in the spatial profiles between circular and linear excitation.
Recently, it has been shown in encapsulated MoSe$_2$ monolayers that the PL intensity at cryogenic temperatures  displays a spatial profile that extends over $1.5 \;\mu$m, for both neutral excitons and trions despite their very different (and short) lifetimes. It has been proposed that the observed PL spatial distribution at low temperatures is likely to be the result of fast hot-exciton  propagation which occurs before relaxing into the light-cone \cite{Park:2021a}. Varying the sample temperature produces a slight change in the measured effective diffusion length $L$, as shown in Fig.\ref{fig:fig2}(c) together with the degree of valley polarization and valley coherence. It is found that $L$ decreases from $0.55\;\mu$m at $T=6$ K down to $0.45\;\mu$m at $T=30$ K. An increase of the scattering rate with temperature can indeed reduce the distance over which hot excitons can travel before energy relaxation.  \\

\noindent
 Another key result which is consistent with our interpretation of the PL profiles is the spatial-dependence of the degree of circular (or linear) polarization of the PL.  As shown in Fig.\ref{fig:fig2}(d)-(e), the polarization is approximately constant in space at low power excitation at $T=6$ K. In our scenario, a spatially-constant polarization is consistent with the fact that the PL spot reflects the initial exciton distribution and not  exciton diffusion. One alternative explanation would be that both the valley polarization and the valley coherence lifetimes are much longer than the exciton lifetime $\tau$, so that no loss of polarization occurs during exciton propagation. If this was the case, however, we should observe a close to $100\;\%$-polarized emission at low temperatures and one should be able to observe a spatial-decay of the polarization at sufficiently high temperature.  Fig. \ref{fig:fig2}(f) shows that changing the sample temperature up to $T=100$ K only changes the overall degree of polarization, but it remains spatially-independent at all temperatures.\\

\begin{figure*}[htbp]
\includegraphics[width=1.05\textwidth]{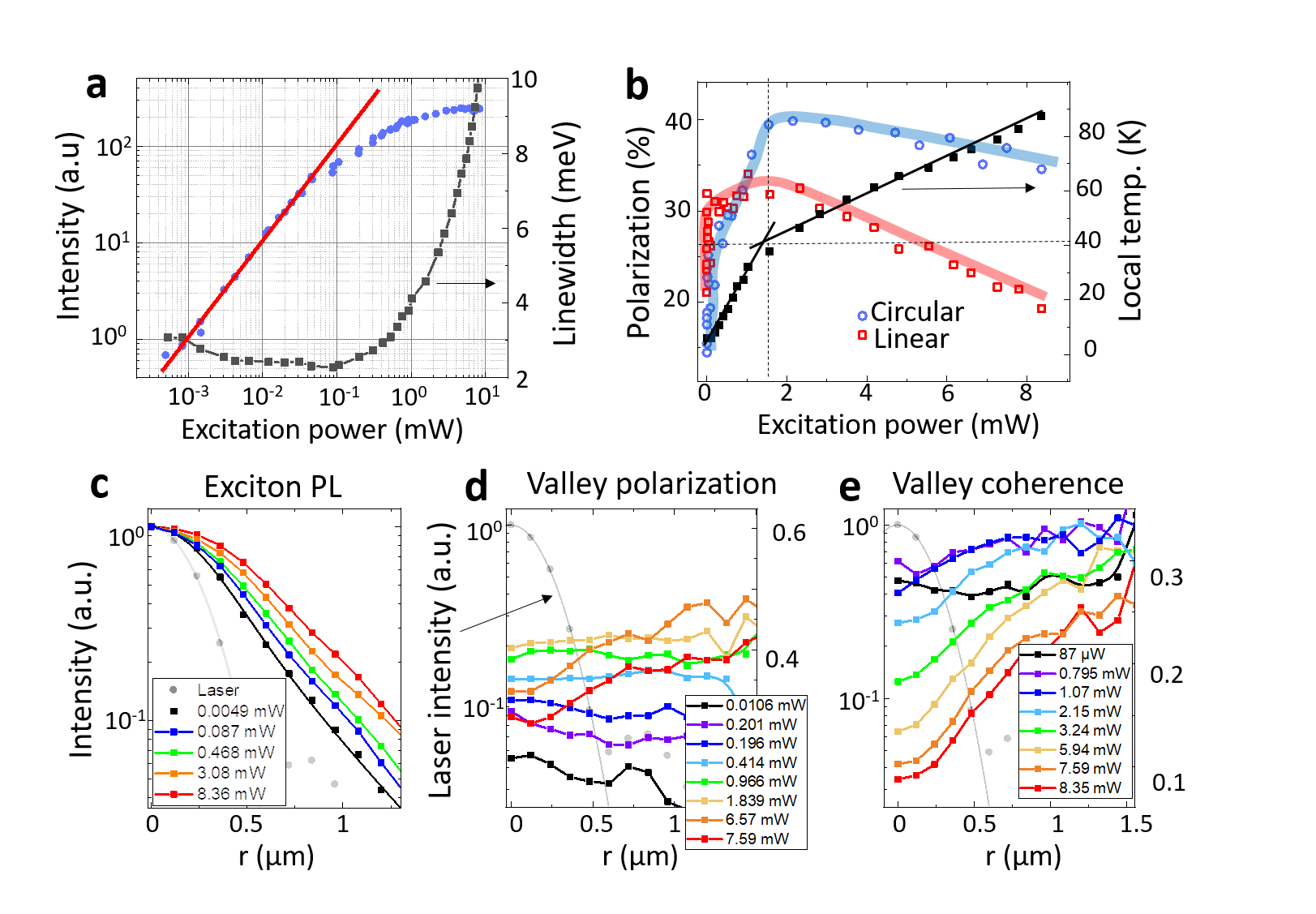}
\caption{\label{fig:fig3} (a) Integrated PL intensity (blue dots) at $T=6$ K as a function of the laser excitation power.  The red line represents a linear relationship between intensity and excitation power. The exciton linewidth is shown in black squares.(b) Degree of polarization as a function of the laser excitation power. Also shown is the linewidth of the neutral exciton peak. The continuous lines are a guide to the eye.
}
\end{figure*}
\noindent
Wu and co-workers \cite{Wu:2021a} have shown that the enhancement of valley polarization can also be achieved by keeping a fixed temperature and adding carriers to the system with the application of a gate bias. This was shown to be detrimental for valley coherence, however, due to an intervalley polaron dressing which results in a higher scattering rate for the in-plane pseudospin. We study another mechanism of valley polarization enhancement: increasing the excitation laser power. Figure \ref{fig:fig3}(a) shows the total PL intensity as a function of the excitation power, in a large range going from 1$\;\mu$W up to 10 mW. Importantly, no change in the PL spectrum is observed after laser exposure at such high power densities, which permits us to exclude the presence of laser-induced photodoping effects  \cite{Cadiz:2016b} thanks to h-BN encapsulation. The linear regime, represented by the red line in  Fig.\ref{fig:fig3}(a), is valid below 100 $\mu$W. Increasing the power leads to a sub-linear behaviour of the exciton luminescence. At 100 $\mu$W, we can roughly estimate the photogenerated exciton density to be $n_0 \sim 2 \times 10^{9}\;\mbox{cm}^{-2}$ by taking an absorption coefficient of $\alpha=1\;\%$, a lifetime $\tau=5$ ps and a uniform distribution inside a circle of radius $L=0.5\;\mu$m. We conclude that
attributing this non-linearity to the onset of Auger-like exciton-exciton recombination would imply an extremely large Auger coefficient of $\gamma\sim 1/(n_0\tau)=100 \; \mbox{cm}^{2}/\mbox{s}$. In addition, the behaviour observed in Fig.\ref{fig:fig3}(a) cannot be described by a simple model based solely on exciton-exciton annihilation since the latter predicts, at high densities, a variation of the intensity $I$ of the form $I\propto \sqrt{P_{ex}}$ where $P_{ex}$ is the excitation power, which does not fit the data.  Instead, we attribute the sub-linear behaviour of the PL intensity to a local heating of the lattice created by the relaxation of hot excitons \cite{Park:2021a}.  At 10 mW, the linewidth increases up to 10 meV (Fig.\ref{fig:fig3}(a)) and the exciton peak redshifts by 5 meV, both consistent with a significant increase of the local temperature up to $T_L \approx 100$ K \cite{Wu:2021a}. This indicates that sources of line broadening other than exciton-phonon interactions do not seem to play a significant role. The PL yield of MoS$_2$ monolayer is reduced by almost one order of magnitude between $6$ K and $100$ K (see supplementary material) and this can significantly contribute to the sub-linear behaviour of the pholuminescence. As the linewidth is much more sensitive to the temperature than the PL emission energy, we have extracted an effective local temperature induced by the laser excitation by comparing the power-induced broadening of the exciton linewidth with the temperature-induced broadening (shown in the supplementary information). \\

\noindent
The result is shown in Fig.\ref{fig:fig3}(b), where two regimes are clearly observed. The energy relaxation rate appears to significantly increase above $T_L=40$ K, similar to what has been observed in bulk III-V semiconductors where the energy relaxation rate was shown to increase with the electron temperature \cite{Ulbrich:1973a}.  Also shown in Fig.\ref{fig:fig3}(b) is the degree of valley polarization and valley coherence as a function of excitation power. We note that $100\;\mu$W also corresponds to the onset of a rapid increase of the valley polarization, with a two-fold increase from $20\;\%$ at low excitation power up to $40\;\%$ achieved at $1.5$ mW.  In this power range, the effective local temperature starts to increase and reaches $T_L=40$ K.   This behaviour is remarkably similar to the temperature-dependence of the valley polarization,  confirming that local laser heating is probably at the origin of the observed enhancement.
Above $T_L=40$ K, valley polarization stops to increase and eventually decreases again, but slightly and slowly. This is different, however, to what is expected based solely on temperature effects, since the polarization should drop sharply above $T_L=40$ K. In addition to the increase of the local temperature, the laser power also modifies the rate of exciton-exciton collisions, which shortens even further the momentum relaxation time $\tau_c$, compensating for the thermal-activation of valley relaxation mechanisms other  than the  MMS mechanism.  This  is similar to what has been observed in GaAs\cite{Oertel:2008a} where the scattering rate increases linearly with the photo-excited electron density and results in an increase of the spin lifetime. Eventually, saturation of the valley polarization will occur, due to screening of interactions, phase space filling and strong local heating effects. 
Note that, in contrast, valley coherence shows a significantly reduced enhancement effect, peaking also at $1.5$ mW after which it starts to rapidly decrease. This is probably due to the  competition between a coherence enhancement due to an increase of the valley relaxation time and a loss of coherence induced by exciton-exciton interactions, expected to be significant at these high photoexcitation densities ($n > n^* \sim 10^{10}\;\mbox{cm}^{-2}$) \cite{Mahmood:2018a}. Indeed, the typical distance between excitons in the  $n=10^{10}-10^{11}\;\mbox{cm}^{-2}$ density range is of the order of $\ell \sim 1/\sqrt{\pi n} = 18-56$ nm, which seems to compare well with the estimated exciton diffusion length. This exciton-exciton interaction is expected to cause similar  detrimental effects on valley coherence as, for example, additional charge carriers \cite{Wu:2021a}.  \\

\noindent
We now focus on the spatial evolution of the valley polarization and coherence when increasing the laser power.
The normalized radial profiles of the exciton luminescence for selected excitation powers are shown in Fig.\ref{fig:fig3}(c). They reveal a moderate broadening of the exciton distribution, probably due to a spatially-dependent PL yield which flattens the radial profiles near $r=0$. It can also be the  beginning of the formation of a halo-like profile due to Seebeck drift under the presence of a temperature gradient \cite{Kulig:2018a,Perea:2019a,Park:2021a}. Remarkably, the radial profiles for the valley polarization and valley coherence exhibit very different behaviour. The valley polarization remains spatially-independent for excitation powers below 2 mW, as shown in Fig.\ref{fig:fig3}(d). This indicates that the mechanism responsible for the valley polarization enhancement in this power range is spatially homogeneous. In our scenario, this implies that the local temperature $T_L$ varies slowly accross the PL spot size. This is consistent with the absence of a clear halo-like profile for the PL intensity \cite{Park:2021a}. At higher powers, exciton-exciton interactions begin to dominate the valley lifetime \cite{Mahmood:2018a} and, therefore, generate a spatial-dependence of the emitted circular polarization which exhibits a small dip at the center above a few mW. Valley coherence, in contrast, develops a significant dip at $r=0$ (Fig.\ref{fig:fig3}(e)) above $1$ mW, consistent with the rapid drop of the  spatially-averaged valley coherence shown in Fig.\ref{fig:fig3}(b).  This suggests that exciton-exciton interactions not only shortens the valley lifetime $\tau_v$, but also significantly increase the pure dephasing rate $\gamma^*$ and therefore doubly affects the valley coherence time $\tau_{vc}$, given by  $1/\tau_{vc}=1/(2\tau_v) + \gamma^*$.  Since both $\tau_{v}$ and $\gamma^*$ vary with the exciton density in this  regime, the measured valley coherence varies rapidly as a function of $r$.\\

In summary, this work brings new elements for the understanding of the different mechanisms that may influence the dynamics of valley polarization and valley coherence in TMD monolayers. We have shown that, in addition to increasing the sample temperature or the resident carrier density, valley polarization can be enhanced by increasing the photogenerated exciton density up to $\sim 10^{10}\;\mbox{cm}^{-2}$. Further increase of the excitation density, together with the significant increase of the local temperature, compensate (and eventually dominate) this enhancement of valley polarization. Valley coherence is shown to be only moderately enhanced, before dropping quickly as a function of the exciton density. This illustrates the detrimental effect of exciton-exciton interactions for the preservation of coherent superposition of valley excitons.\\

\indent \textit{Acknowledgements.---} 
F.C, acknowledges the Grant "SpinCAT" No. ANR-18-CE24-0011-01.  F.C. would like to thank X. Marie and C. Robert for fruitful discussions. K.W. and T.T. acknowledge support from the Elemental Strategy Initiative
conducted by the MEXT, Japan (Grant Number JPMXP0112101001) and  JSPS
KAKENHI (Grant Numbers 19H05790, 20H00354 and 21H05233).

%

\end{document}